\documentclass[aps,pra,preprint,groupedaddress,showpacs]{revtex4}

\usepackage{graphicx}% Include figure files
\usepackage{bm}% bold math

\begin{document}

%\preprint{}

\title{R-matrix calculation of electron collisions
with electronically excited O$_2$ molecules}


\author{Motomichi Tashiro}
\email[]{tashiro@euch4e.chem.emory.edu}
\author{Keiji Morokuma}
%\thanks{}
\affiliation{Department of Chemistry, Emory University, 1515 Dickey Drive, Atlanta, Georgia 30322, USA.}

\author{Jonathan Tennyson}
\affiliation{Department of Physics and Astronomy,
University College London, London WC1E 6BT, UK.}


\date{\today}

\begin{abstract}
Low-energy electron collisions with O$_2$ molecules are 
studied using the fixed-bond R-matrix method. In addition 
to the O$_2$ ${X}^3\Sigma_{g}^-$ ground state, integrated cross 
sections are calculated for elecron collisions with 
the ${a}^1\Delta_{g}$ and 
${b}^1\Sigma_{g}^+$ excited states of O$_2$ molecules. 
13 target electronic states of O$_2$ are included 
in the model within a valence configuration interaction 
representations of the target states. 
Elastic cross sections for the ${a}^1\Delta_{g}$ and 
${b}^1\Sigma_{g}^+$ excited states are similar to the cross 
sections for the ${X}^3\Sigma_{g}^-$ ground state. 
As in case of excitation from the ${X}^3\Sigma_{g}^-$ state, 
the O$_2^-$ $\Pi_u$ resonance makes the dominant contribution 
to excitation cross sections from the ${a}^1\Delta_{g}$ and 
${b}^1\Sigma_{g}^+$ states. 
The magnitude of excitation cross sections from the ${a}^1\Delta_{g}$
state to the ${b}^1\Sigma_{g}^+$ state is about 10 time larger than
the corresponding cross sections from the ${X}^3\Sigma_{g}^-$ to 
the ${b}^1\Sigma_{g}^+$ state. For this ${a}^1\Delta_{g}$ $\to$ 
${b}^1\Sigma_{g}^+$ transition, our cross section at 4.5 eV agrees 
well with the available experimental value. 
These results should be important for models of plasma discharge 
chemistry which often requires cross 
sections between the excited electronic states of O$_2$. 
\end{abstract}

%\keywords{}

\maketitle

%\section{}
%\subsection{}
%\subsubsection{}

\section{Introduction}
%original motivation/situation of exp.results
%--> we need to check exp.data for references

%motivation
An understanding of electron collision processes with oxygen molecules
is important because of its role in chemistry of electrical discharge 
and the upper atmosphere. 
In recent attempts to an operate electrical discharge oxygen-iodine laser, 
a population inversion of iodine atoms was achieved by 
a near resonant energy transfer via the 
${\rm O}_2({a}^1\Delta_{g})+{\rm I}({}^2P_{3/2})$ 
$\leftrightarrow$ ${\rm O}_2({X}^3\Sigma^{-}_{g})+{\rm I}({}^2P_{1/2})$ 
process. 
In contrast to the traditional liquid chemistry singlet oxygen
generator \cite{Mc78}, recent studies \cite{Ca05a,Ca05b} utilize
flowing electric discharges where electron collisions with O$_2$ 
excited electronic states can be important \cite{Sh96,Gu04}. 
In such conditions, even highly excited metastable states of 
O$_2$(${c}^1\Sigma^{-}_{u}$, ${A'}^3\Delta_{u}$,
${A}^3\Sigma^{+}_{u}$) may play roles \cite{Ha99}, 
in addition to the lower O$_2$ ${a}^1\Delta_{g}$ and
${b}^1\Sigma_{g}^+$ excited states. 

However, most previous work has concentrated on electron
collisions with the O$_2({X}^3\Sigma^{-}_{g})$ ground state, 
so our knowledge of electron impact transitions from the 
O$_2 {a}^1\Delta_{g}$ and ${b}^1\Sigma_{g}^+$ excited states 
is limited. 
The past experimental and theoretical works concerning electron 
O$_2$(${X}^3\Sigma^{-}_{g}$, ${a}^1\Delta_{g}$) collisions were 
summarized by Brunger and Buckman \cite{Br02}. 
One work on the excited electronic states is measurement of 
differential and integral cross sections at 4.5 eV for excitation from 
the O$_2$ ${a}^1\Delta_{g}$ state to the ${b}^1\Sigma_{g}^+$ state by 
Hall and Trajmar \cite{Ha75}. Their value is more than an order of 
magnitude larger than that for the ${X}^3\Sigma^{-}_{g}$ $\rightarrow$ 
${b}^1\Sigma_{g}^+$ cross section. 
Also, Khakoo et al. \cite{Kh83} studied the energy-loss spectrum
for electron impact excitation on discharged O$_2$ and assigned 
the transitions O$_2$ ${a}^1\Delta_{g}$ ($v$=0) 
$\rightarrow$ O$_2$ ${}^1\Pi_{u}$ ($v'$=0,1,..7). 
Burrow \cite{Bu73} and Beli\'c and Hall \cite{Be81}
studied dissociative electron attachment with the O$_2$
${a}^1\Delta_{g}$ state. 
The later authors found that dissociation proceeds to 3 different 
limits, O${}^{-}({}^2P)$+O$({}^3P)$, O${}^{-}({}^2P)$+O$({}^1D)$ 
and O${}^{-}({}^2P)$+O$({}^1S)$. 

In contrast to the situation in electron collisions with the excited  
oxygen molecule, a lot of work has been performed on the ground state 
O$_2$, both experimental \cite{Te87,Mi92,Mi94,Wo95,Wo96} 
and theoretical 
\cite{Te87,Mi94,Wo95,Wo96,No86,No92,Hi94,Hi95,No96}.
Notably, Noble, Burke and their co-workers extensively applied 
their R-matrix method to the electron O$_2$ collision problems during 
1992-1996 \cite{Mi94,Wo95,Wo96,No92,Hi94,Hi95,No96}. 
They studied electronic excitation processes from the 
O$_2({X}^3\Sigma^{-}_{g})$ ground state to the ${a}^1\Delta_{g}$, 
${b}^1\Sigma_{g}^+$, ${c}^1\Sigma^{-}_{u}$, ${A'}^3\Delta_{u}$ and 
${A}^3\Sigma^{+}_{u}$ states using the fix-bond R-matrix 
method \cite{No92,Hi94} and applied the non-adiabatic R-matrix method
to vibrational excitations process of the ${X}^3\Sigma^{-}_{g}(v=0)$ 
$\rightarrow$ ${X}^3\Sigma^{-}_{g}(v=0-4)$ transitions 
\cite{Hi95,No96}. 
They also calculated differential cross sections for elastic electron 
collisions of the ${X}^3\Sigma^{-}_{g}$ state \cite{Wo95} as well as 
impact excitations from the ${X}^3\Sigma^{-}_{g}$ state to 
the ${a}^1\Delta_{g}$ and ${b}^1\Sigma_{g}^+$ states \cite{Mi94}. 
The effect of nuclear motion was included in the former elastic cross 
sections by vibrational averaging of the T-matrix \cite{Wo95}. 
Other than these R-matrix calculations, Teillet-Billy et al. \cite{Te87}
applied effective range theory (ERT) to excitations from the 
${X}^3\Sigma^{-}_{g}$ to the ${a}^1\Delta_{g}$ and ${b}^1\Sigma_{g}^+$ 
states. Because of the different treatment of the O$_2^{-}$ resonances, 
the ERT results deviate from the R-matrix cross sections 
at energies above 5 eV. 

Given the importance of electron collisions with excited 
O$_2$ molecules, we perform R-matrix calculations for electron 
O$_2$(${a}^1\Delta_{g}$,${b}^1\Sigma_{g}^+$) collisions. 
We chose the R-matrix method because it has been successfully applied
to many electron-molecule collisions including e-N$_2$, N$_2$O and 
H$_2$O \cite{Te99,Bu05,Go05}. 
The fixed-bond method was employed in this work, because it gave
reasonably good results in previous studies \cite{No86,No92,Hi94} 
for transitions from the O$_2$ ${X}^3\Sigma^{-}_{g}$ state to 
the ${a}^1\Delta_{g}$, ${b}^1\Sigma_{g}^+$ state, and the `6 eV states' 
(${c}^1\Sigma^{-}_{u}$+${A'}^3\Delta_{u}$+${A}^3\Sigma^{+}_{u}$). 
In addition to these 6 low lying O$_2$ target states, 
previous calculations included three higher excited target 
states of O$_2$ ${B}^3\Sigma^{-}_{u}$,${1}^1\Delta_{u}$ and
${f'}^1\Sigma^{+}_{u}$, in order to improve quality of 
the R-matrix calculations \cite{No92,Hi94}.  
In this work, we use a valence complete active space description of 
the O$_2$ target states and add other valence target states, 
${1}^1\Pi_{g}$,${1}^3\Pi_{g}$,${1}^1\Pi_{u}$,${1}^3\Pi_{u}$, in our 
calculations. Since excitation energies of some of these states are 
lower than those of ${B}^3\Sigma^{-}_{u}$,${1}^1\Delta_{u}$ and 
${f'}^1\Sigma^{+}_{u}$ states, some improvement can be expected by 
inclusion of these extra $\Pi$ target states. 
In principle, a complete valence active space is not sufficient for 
the description of these targets, because some of them are mixed with 
n=3 Rydberg states as described in Buenker and Peyerimhoff 
\cite{Bu75a,Bu75b}. 
Since expansion of this active space increases the calculation cost 
considerably, we limit ourselves here the inclusion of the valence 
states to test the effects of the higher excited states. 

In this paper, details of the calculations are presented in section 2, 
and we discuss the results in section 3 comparing our results with 
previous theoretical and available experiments. 
Then the summary is given in section 4. 

\clearpage

\section{Theoretical methods}

The R-matrix method has been described extensively in the previous 
literature \cite{Bu05,Go05,Mo98}, 
so here we only repeat the outline of the method. 
In this method, configuration space is divided by two regions
according to the distance $r_{N+1}$ of the scattering electron and  
the center of mass of the target molecule having $N$ electrons. 
In the inner region $r_{N+1} < a$, the $N$+1 electrons problem 
is solved by usual quantum chemistry method with slight modifications  
to account for existence of boundary at $r_{N+1}=a$. 
In the inner region, the total $N$+1 electrons wave functions 
are represented by $N$-electron CI target wave functions augmented 
by diffuse functions. 
Here the target wave functions are contained in the sphere 
$r_{N+1} < a$, whereas the diffuse functions overlap the 
boundary at $r_{N+1} = a$ in order to describe the scattering electron. 
In the outer region $r_{N+1} > a$, the problem is reduced to single 
electron scattering, ignoring exchange of the scattering electron with 
the target electrons. Interaction of the scattering electron and the
target is considered through static multipolar interaction terms which 
introduce inter-channel couplings. 
The wave functions obtained in the inner region are converted to the 
R-matrix at the boundary $r_{N+1}$=$a$, then the coupled radial
Schr\"odinger equations are solved so as to extract scattering
information at the asymptotic region.

In the inner region, the $N$+1 electronic wavefunctions are expanded
as, 
\begin{equation}
\Psi = \textrm{$\mathcal{A}$} \sum_{ij} \Phi_{i} \left(1...N;R \right) 
u_{j} \left(N+1;R \right) a_{ij} 
+ \sum_{q} X_{q} \left(1...N+1;R \right) b_{q}, 
\label{}
\end{equation}
where $\mathcal{A}$ is an antisymmetrization operator,  
$\Phi_{i}$ are the $N$ electron target CI wave functions, 
$u_{j}$ are the diffuse functions representing wave functions of a 
scattering electron,
and $X_{q}$ are bound $N$+1 electron wave functions, 
while $a_{ij}$ and $b_{q}$ are variational coefficients. 
In this expression, the first term represents the scattering of an 
electron from and to the asymptotic region. 
The second summation involves purely 
$L^2$ integrable terms. In addition to the target molecular orbitals 
included in the CI wavefunctions in the first summation, 
some extra target virtual orbitals are usually included in $X_{q}$ 
in order to account for short range polarization effects. 

\begin{figure}
 \includegraphics{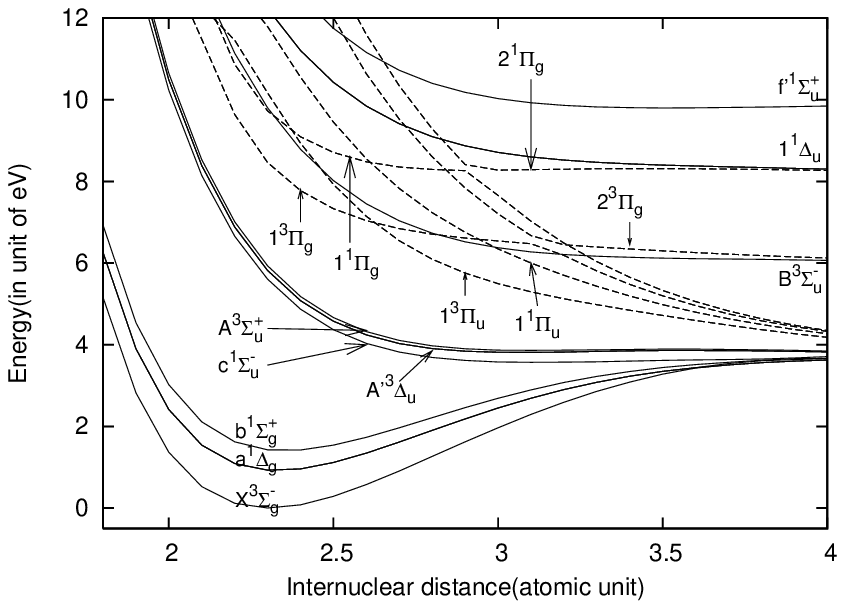}%
 \caption{\label{fig1} 
   Potential energy curves of the O$_2$ electronic states.
   The equilibrium distance of the ${X}^3\Sigma^{-}_{g}$ state, 
   $R$ = 2.3 a$_0$ is used in our R-matrix calculations.  
   }
\end{figure}

We used a modified version of the polyatomic programs in the UK
molecular R-matrix codes \cite{Mo98}.  
These programs utilize gaussian type orbitals (GTO) to represent 
target molecule as well as a scattering electron. 
Although most of past R-matrix works on electron O$_2$ collisions 
had employed the diatomic modules using Slater type orbitals (STO) 
obtained by Hartree Fock(HF) calculation, 
we select GTO mainly because of simplicity of the input and 
availability of basis functions. 
The state averaged complete active space SCF (SA-CASSCF) orbitals 
are imported from the target calculations with MOLPRO suites of 
programs \cite{molpro}. 
This employment of SA-CASSCF orbitals improves the vertical 
excitation energies of the O$_2$ target states compared to the 
energies obtained using HF orbitals. 
These target orbitals are constructed from the [5s,3p] contracted
basis of Dunning \cite{Du71} augmented by a d function with exponent 
1.8846, as in Sarpal et al. \cite{Sa96}. 
In the R-matrix calculations, we included 13 target states; 
${X}^3\Sigma^{-}_{g}$,${a}^1\Delta_{g}$,
${b}^1\Sigma^{+}_{g}$,${c}^1\Sigma^{-}_{u}$,${A'}^3\Delta_{u}$,
${A}^3\Sigma^{+}_{u}$,${B}^3\Sigma^{-}_{u}$,${1}^1\Delta_{u}$,
${f'}^1\Sigma^{+}_{u}$, 
${1}^1\Pi_{g}$,${1}^3\Pi_{g}$,${1}^1\Pi_{u}$ and ${1}^3\Pi_{u}$, 
where the last 4 $\Pi$ states were not included in previous
calculations. 
The potential energy curves of these target electronic states are
shown in figure \ref{fig1} for reference. 
Further details of these target electronic states can be found in 
Saxon and Liu \cite{Sa77} and Minaev and Minaeva \cite{Mi01} for
example. 
In our fixed-bond R-matrix calculations, these target states are
evaluated at the equilibrium bond length $R$ = 2.3 a$_0$ 
of the O$_2$ ${X}^3\Sigma^{-}_{g}$ ground electronic state. 
Note that all calculations were performed with D$_{\rm 2h}$ symmetry
because of restriction of the polyatomic UK R-matrix codes,  
though natural symmetry of this system is D$_{\infty h}$. 

The radius of the R-matrix sphere $a$ was chosen to be 10 a$_0$ in our  
calculations. 
In order to represent the scattering electron, we included diffuse
gaussian functions up to $l$=5 with 9 functions for $l$=0, 7 functions 
for $l$=1-3 and 6 functions for $l$=4 and 5.  
The exponents of these gaussians were fitted using the GTOBAS 
program \cite{Fa02} in the UK R-matrix codes.  
Details of the fitting procedure are the same as in Faure 
et al. \cite{Fa02}. 
We constructed the $N+1$ electron configurations from the orbitals
listed in table \ref{tab0}. 
The CI target wave functions are composed from the valence orbitals in 
table \ref{tab0} with the 1$a_g$ and 1$b_{1u}$ orbitals kept doubly 
occupied. 
The first terms in equation (1) are constructed from configurations 
of the form, 
\begin{equation}
1a_g^2 1b_{1u}^2 \{ 2a_g 3 a_g 1 b_{2u} 1 b_{3u} 2 b_{1u} 3 b_{1u} 
1 b_{3g} 1 b_{2g} \}^{12}  
\left( {}^{3} B_{1g} \right) \{2b_{1g}...17b_{1g} \}^{1} 
\left( {}^{2}A_g \right), 
\end{equation}
here we assume that the total symmetry of this 17 electrons system is
${}^2A_g$. 
The first 4 electrons are always kept in the 1$a_g$ and 1$b_{1u}$
orbitals, then the next 12 electrons are distributed over the valence
orbitals with restriction of target state symmetry, ${}^{3} B_{1g}$
symmetry of the O$_2$ ground state in this case. 
The last electron, the scattering electron, occupies one of the
diffuse orbitals, $B_{1g}$ symmetry in this example. 
To complete the wave function with the total symmetry ${}^2A_g$, 
we also have to include configurations with the other target states 
combined with diffuse orbitals having appropriate symmetry in the same
way as in the example. 
The second terms in equation (1) are constructed from configurations,  
\begin{equation} 
1a_g^2 1b_{1u}^2 \{ 2a_g 3 a_g 1 b_{2u} 1 b_{3u} 2 b_{1u} 3 b_{1u} 
1 b_{3g} 1 b_{2g} \}^{12}  
\left( {}^{3} B_{1g} \right) \{ 1b_{1g} \}^{1} \left( {}^{2}A_g
\right),
\end{equation}
where the scattering electron occupies a bound $1b_{1g}$ extra virtual 
orbital, instead of the diffuse continuum orbitals in the 
expression (2). 
As in table \ref{tab0}, we included one extra virtual orbital for each
symmetry. 
The second terms in equation (1) also contain configurations of the form  
\begin{equation}
1a_g^2 1b_{1u}^2 \{ 2a_g 3 a_g 1 b_{2u} 1 b_{3u} 2 b_{1u} 3 b_{1u} 1
b_{3g} 1 b_{2g} \}^{13} 
\left( {}^{2}A_g \right).
\end{equation} 
In this case, the last 13 electrons including the scattering electron 
are distributed over the valence orbitals with the restriction of 
${}^2A_g$ symmetry. 
In this way, the number of configurations generated for a specific total 
symmetry is typically about 17000, though the final dimension of the inner 
region Hamiltonian 
is reduced to be about 500 by using CI target contraction and 
prototype CI expansion method \cite{Te95}. 

In order to obtain the integral cross sections for the electron O$_2$ 
collisions, the R-matrix calculations were performed over all 8
irreducible representations of D$_{2h}$ symmetry, 
$A_g$,$B_{2u}$,$B_{3u}$,$B_{1g}$,$B_{1u}$,$B_{3g}$,$B_{2g}$ and $A_u$ 
with both doublet and quartet spin multiplicity.

\clearpage

\section{Results and discussion}

Figure \ref{fig1} shows the potential energy curves of the O$_2$ target
states. 
These curves were calculated by the SA-CASSCF method which was 
used in the actual R-matrix calculations. Although not included 
in our R-matrix calculations, we also include the curves for the 
O$_2({2}^{1,3}\Pi_g)$ states for reference. 
Table \ref{tab1} compares vertical excitation energies from the present 
calculations with previous HF/STO results. 
Compared to the experimentally estimated values, our results 
are of the same quality for the O$_2$ ${a}^1\Delta_{g}$ state and are 
slightly worse, 0.05 eV, than the HF/STO result for the 
${b}^1\Sigma^{+}_{g}$ state. However, the excitation energies are
improved by about 0.1 eV for the `6 eV states'; 
${c}^1\Sigma^{-}_{u}$, ${A'}^3\Delta_{u}$ and 
${A}^3\Sigma^{+}_{u}$. 
For the higher 3 electronic states, the improvement is about 1 eV. 
Though discrepancies with the experimental values are still not small, 
0.22 eV for the ${b}^1\Sigma^{+}_{g}$ state and 0.45 eV for 
the ${A'}^3\Delta_{u}$ state for examples, 
we believe that our choice of the GTO basis set and the 
CAS space is satisfactory for the present R-matrix calculations 
considering the differences with the previous HF/STO results. 

\begin{figure}
 \includegraphics{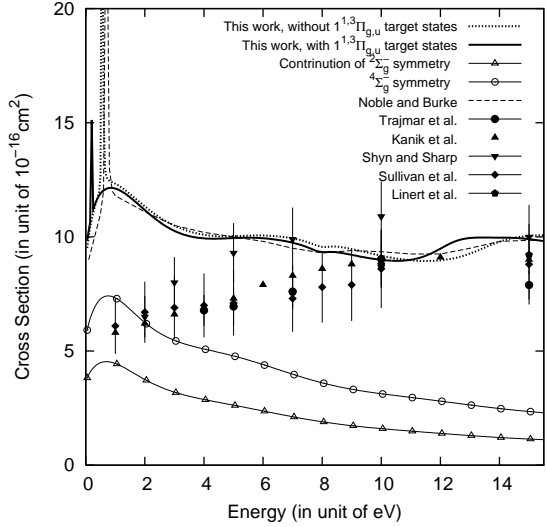}%
 \caption{\label{fig2}
  The elastic cross sections of the O$_2$ ${X}^3\Sigma^{-}_{g}$ state. 
  Thick Full line represents cross sections obtained by 13 target states 
  calculation including ${1}^{1,3}\Pi_{g,u}$ target states. 
  Thick dotted line is the cross sections including 9 target states 
  without ${1}^{1,3}\Pi_{g,u}$ target states. 
  The partial cross sections from the 13 target states calculation are 
  represented 
  by thin full lines marked with open symbols. Symmetries 
  with minor contributions are not shown in the figure.  
  For comparisons, we also include the previous R-matrix results of 
  Noble and Burke \cite{No92}, the experimental cross sections of 
  Trajmar et al. \cite{Tr71}, Kanik et al. \cite{Ka93}, Shyn and 
  Sharp \cite{Sh82}, Sullivan et al. \cite{Su95} and 
  Linert et al. \cite{Li04}. 
  }
\end{figure}

\begin{figure}
 \includegraphics{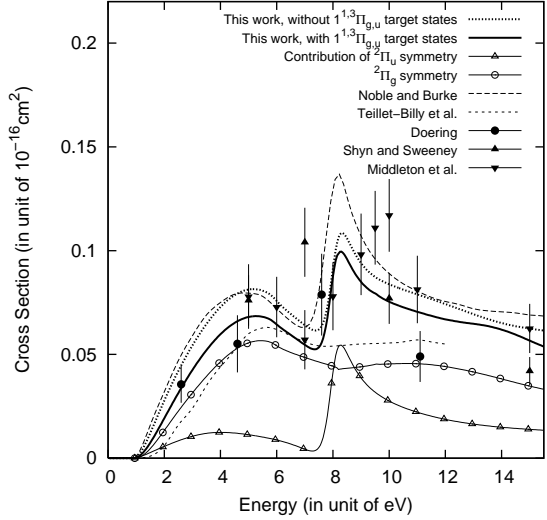}%
 \caption{\label{fig3}
  The excitation cross section from the O$_2$ ${X}^3\Sigma^{-}_{g}$ 
  state to the ${a}^1\Delta_{g}$ state.
  Our results are shown in thick full and dotted lines as in 
  figure \ref{fig2}. 
  The partial cross sections of ${}^2 \Pi_{g,u}$ symmetries are 
  also shown as thin full lines marked with open symbols. 
  For comparison, we include the previous R-matrix results of 
  Noble and Burke \cite{No92}, the ERT calculations of Teillet-Billy 
  et al. \cite{Te87}, the experimental cross sections of 
  Doering \cite{Do92}, Shyn and Sweeney \cite{Sh93} 
  and Middleton et al. \cite{Mi92}. 
  }
\end{figure}

\begin{figure}
 \includegraphics{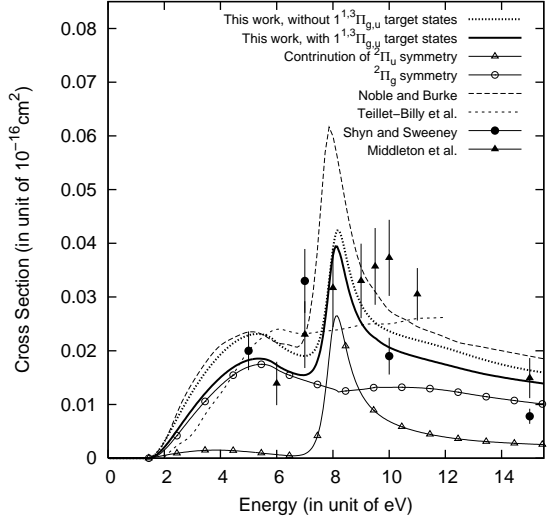}%
 \caption{\label{fig4}
  The excitation cross section from the O$_2$ ${X}^3\Sigma^{-}_{g}$ 
  state to the ${b}^1\Sigma^{+}_{g}$ state.
  Our results are shown in thick full and dotted lines as in 
  figure \ref{fig2}.   
  We include the previous theoretical results of 
  Noble and Burke \cite{No92}, Teillet-Billy et al. \cite{Te87} and 
  the experimental cross sections of Shyn and Sweeney \cite{Sh93} 
  and Middleton et al. \cite{Mi92}. 
  }
\end{figure}

\begin{figure}
 \includegraphics{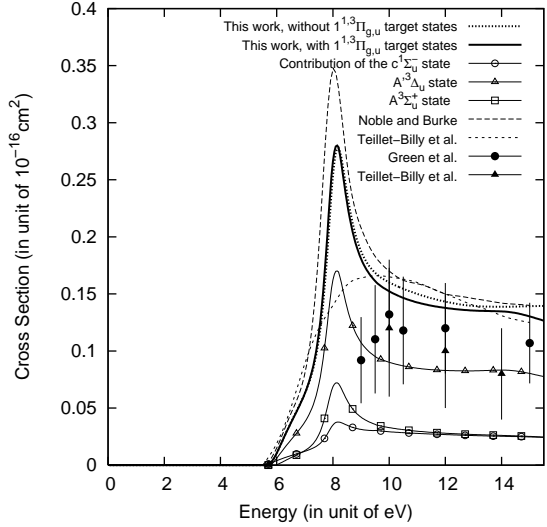}%
  \caption{\label{fig5}
  The excitation cross sections from the O$_2$ ${X}^3\Sigma^{-}_{g}$ 
  state to the `6 eV states' which consist of the O$_2$ 
  ${c}^1\Sigma^{-}_{u}$,${A'}^3\Delta_{u}$ and 
  ${A}^3\Sigma^{+}_{u}$ states. The total cross sections shown here 
  are the sum of the individual cross sections of these 3 states. 
  Our results for the total cross sections are shown in thick full and 
  dotted lines as in figure \ref{fig2}. 
  The individual excitation cross sections from 13 states calculations 
  are shown as thin full lines marked with open symbols. 
  We include the total cross sections from the previous R-matrix 
  results of Noble and Burke \cite{No92}, 
  the ERT calculations of Gauyacq et al. \cite{Ga88}, the experimental 
  results of Teillet-Billy et al. \cite{Te89} and 
  Green et al. \cite{Gr01}. 
  }
\end{figure}

In figures \ref{fig2}-\ref{fig5}, the cross sections 
for the transitions from the O$_2$(${X}^3\Sigma^{-}_{g}$) ground state 
are shown. 
These cross sections were previously calculated using the R-matrix method, 
but with a different basis set and target descriptions \cite{No86,No92}. 
In figure \ref{fig2}, we compare our elastic scattering cross sections 
for the ${X}^3\Sigma^{-}_{g}$ state 
with the previous theoretical and experimental results.  
The theoretical elastic scattering cross sections are quite similar in 
shape and magnitude each other. 
There is a sharp peak around 0.5 eV in each theoretical cross
section, which comes from the O$_2^{-}$ ${}^{2} \Pi_g$ resonance. 
In our calculation with 13 target states including 4 extra $\Pi$
targets, this ${}^{2} \Pi_g$ resonance is located at 0.196 eV and the
width is 0.00134 eV. 
When the number of targets are reduced to 9 by removing the $\Pi$
target states, the location of the resonance is shifted to 0.548 eV
with a width of 0.0161 eV. 
The later resonance parameters with 9 target states are closer to 
the results of the previous calculations, reflecting inclusion of the
same number of the target states.  
When our results are compared to the experimentally measured elastic 
cross sections, agreement is good for energy above 10 eV but is  
poorer at lower scattering energy below 5 eV. 
At 1 eV, the theoretical cross section is about a factor of 2 
larger than the experimental results. 
This situation mirrors that in the previous R-matrix calculations. 
As discussed in Noble and Burke \cite{No92}, 
this discrepancy is attributed to lack of long-range polarization 
effects in our and their model. To improve this low energy 
behaviour of the elastic cross sections, we may need 
pseudostates method of Gillan et al. \cite{Gi88} and 
Gorfinkiel and Tennyson \cite{Go05b} for example. 

%X->a and b 
Figures \ref{fig3} and \ref{fig4} show excitation cross 
sections from the O$_2$ ${X}^3\Sigma^{-}_{g}$ state 
to the ${a}^1\Delta_{g}$ and ${b}^1\Sigma^{+}_{g}$
states. 
In both figures, there is a pronounced peak in the cross sections
around 8 eV which comes from the O$_2^{-}$ ${}^2 \Pi_u$ resonance
located at 7.988 eV with its width being 0.906 eV. 
Compared to the previous R-matrix calculations of 
Noble and Burke \cite{No92}, our cross sections with 13 target states 
are slightly smaller at all scattering energies.  
The peak height of our results around 8 eV is 30\% smaller 
in excitation to the ${a}^1\Delta_{g}$ case and is 35\% smaller in 
the ${b}^1\Sigma^{+}_{g}$ case. 
However, general feature of the cross section profiles are quite
similar in our results and the previous R-matrix calculations. 
We also compare our results with the effective range theory (ERT) 
results of Teillet-Billy et al. \cite{Te87}. 
Their method relied on the existence of the O$_2^{-}$ ${}^{2} \Pi_g$ 
resonance around 0.2 eV, but did not include the effect of the 
O$_2^{-}$ ${}^{2} \Pi_u$ resonance located at 8 eV. 
Thus, their results and our cross sections agree well at low energy, 
below 6 eV, where the ${}^{2} \Pi_g$ symmetry mainly contributes 
to the total cross sections. 
However, the agreement is worse at energy range above 7 eV because 
of the ${}^{2} \Pi_u$ resonance contributions. 
In both O$_2$(${a}^1\Delta_{g}$) and O$_2$(${b}^1\Sigma^{+}_{g}$)
cases, agreement with the experimental cross sections is modestly good 
in the energy regions away from the resonance peak. 
As shown in the figures \ref{fig3} and \ref{fig4}, the cross sections
of Middleton et al. \cite{Mi92} have peak in energy region around 
10 eV, which is 2 eV larger than the theoretical position. 
To resolve this discrepancy, we need to include the effect of nuclear 
motion in calculation as discussed in Higgins et al. \cite{Hi94}. 

%X->6eV
Figure \ref{fig5} shows excitation cross sections 
from the O$_2$ ${X}^3\Sigma^{-}_{g}$ state to the `6 eV states'. 
These `6 eV states' consist of the O$_2$ ${c}^1\Sigma^{-}_{u}$, 
${A'}^3\Delta_{u}$ and ${A}^3\Sigma^{+}_{u}$ states. 
In order to compare with previous experimental measurements, 
we sum the cross sections for the transitions to these states. 
As in the case of transitions to the ${a}^1\Delta_{g}$ and 
${b}^1\Sigma^{+}_{g}$ states in figures \ref{fig3} and \ref{fig4}, 
a prominent peak exists around 8 eV in our results. 
The cause of this peak is the O$_2^{-}$ ${}^{2} \Pi_u$ resonance 
located at 7.988 eV. 
Our results are quite similar in shape and magnitude to the previous 
R-matrix calculation of Noble and Burke \cite{No92}, though the cross 
section peak at 8 eV is slightly lower in our case.  
The ERT results of Gauyacq et al. \cite{Ga88} are also shown in the 
figure. 
As discussed above, their cross sections do not have a peak around 
8 eV because they did not include the O$_2^{-}$ ${}^{2} \Pi_u$ 
resonance effects. 
Recently, Green et al.\cite{Gr01} measured the integral cross sections 
from the ${X}^3\Sigma^{-}_{g}$ state to the `6 eV states' and
discussed the discrepancy between the past theoretical results and
their measurements. 
Though the theoretical cross sections have a peak around 8 eV,  
the experimental results do not show this peak nor the enhancement of 
the cross sections near 8 eV. 
Figure \ref{fig5} compares our results with the experimental
cross sections, which shows that the discrepancy still exists below 
10 eV. 
This deviation may come from our use of fixed-bond approximation, 
because the equilibrium bond distances of 
`6 eV state' are longer than those of ${X}^3\Sigma^{-}_{g}$,
${a}^1\Delta_{g}$ and ${b}^1\Sigma^{+}_{g}$ states.  
In principle, we need to employ the non-adiabatic R-matrix method or 
vibrational averaging procedure to take into account those 
difference of the equilibrium distances. In this study, we limit 
ourselves at the fixed-bond approximation 
and leave the treatment of nuclear motion for work in future. 

\begin{figure}
 \includegraphics{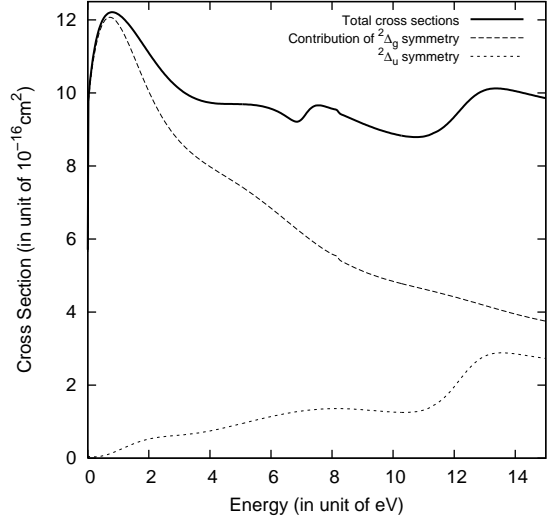}%
 \caption{\label{fig6}
The elastic cross sections for the O$_2$ ${a}^1\Delta_{g}$ state. 
The thick full line represents the cross sections obtained by 13
target states calculation including ${1}^{1,3}\Pi_{g,u}$ target 
states. The contributions of the ${}^{2}\Delta_{g}$ and 
${}^{2}\Delta_{u}$ total symmetries are also shown.   
 }
\end{figure}

Figure \ref{fig6} shows elastic cross sections for the ${a}^1\Delta_{g}$ 
state as functions of electron collision energy. 
These cross sections have almost the same shape and magnitude as the 
${X}^3\Sigma^{-}_{g}$ state elastic cross sections. 
We do not observe a sharp resonance peak in the ${a}^1\Delta_{g}$ 
elastic cross sections in contrast to the ${X}^3\Sigma^{-}_{g}$ case,   
because the O$_2^{-}$ ${}^{2} \Pi_g$ resonance is located 0.7 eV below 
the O$_2$ ${a}^1\Delta_{g}$ state, but 0.2 eV above the 
O$_2$ ${X}^3\Sigma^{-}_{g}$ state. As shown in the figure \ref{fig6}, 
the ${}^2\Delta_{g}$ symmetry is the main contributor to the 
cross sections at low energy. 
This indicates that the $l$=0 component of the 
scattering electron is as important as for the ${X}^3\Sigma^{-}_{g}$ 
elastic scattering. 

\begin{figure}
  \includegraphics{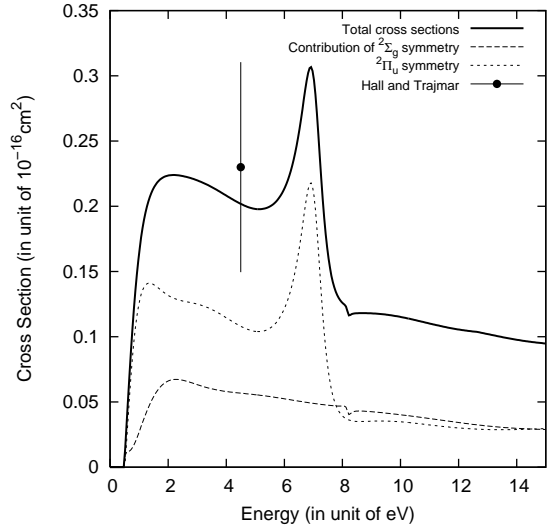}%
  \caption{\label{fig7}
Excitation cross section from the O$_2$ ${a}^1\Delta_{g}$ state 
to the ${b}^1\Sigma^{+}_{g}$ state. 
Our results are shown in thick full line with the contributions 
of ${}^{2}\Sigma_{g}^+$ and ${}^{2}\Pi_{u}$ symmetries in thin lines 
as in figure \ref{fig6}.  
The experimental cross section of Hall and Trajmar \cite{Ha75} is 
also included for comparison. 
 }
\end{figure}

%a->b
The cross section for excitation to the ${b}^1\Sigma^{+}_{g}$
state from the ${a}^1\Delta_{g}$ state is shown in 
figure \ref{fig7}. 
The magnitude of the cross section is about 10 times larger 
than the corresponding cross section for excitation from the 
${X}^3\Sigma^{-}_{g}$ state to the ${b}^1\Sigma^{+}_{g}$ state. 
At electron collision energy of 7.0 eV, there is a large peak 
in the cross sections arising from the ${}^2\Pi_{u}$ symmetry. 
The origin of this peak is the O$_2^{-}$ ${}^{2} \Pi_u$ resonance 
as in the cross sections from the ${X}^3\Sigma^{-}_{g}$ state shown 
in figures \ref{fig3}-\ref{fig5}. 
Because the cross sections are plotted as functions of electron 
collision energy, the positions of the peak in figure \ref{fig7} and 
figure \ref{fig3}-\ref{fig5} are different by 0.93 eV which is the 
energy difference of the O$_2$ ${X}^3\Sigma^{-}_{g}$ state and the 
O$_2$ ${a}^1\Delta_{g}$ state. 
Hall and Trajmar experimentally determined differential and integral 
cross sections at 4.5 eV for this excitation \cite{Ha75}. 
As in figure \ref{fig7}, their value and our results agree well.   
However, we need more experimental data to make detailed comparisons. 

\begin{figure}
  \includegraphics{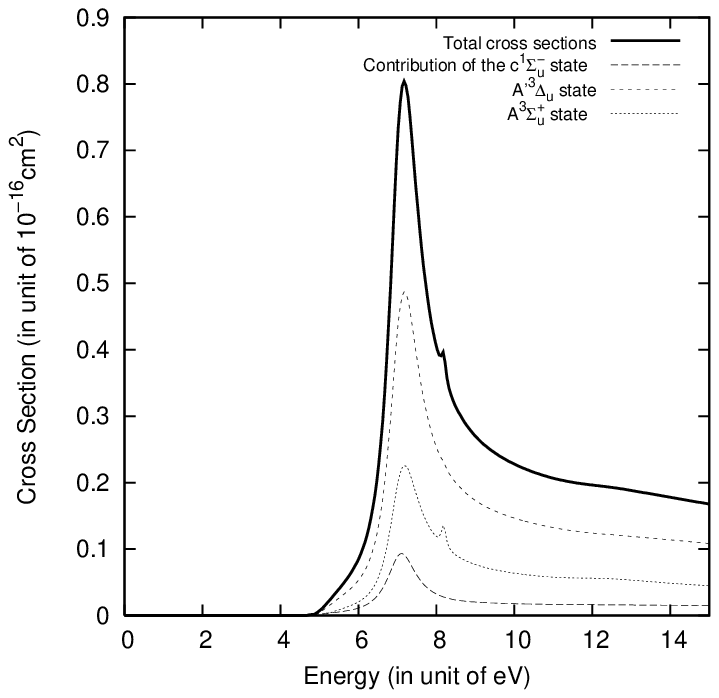}%
  \caption{\label{fig8}
Excitation cross sections from the O$_2$ ${a}^1\Delta_{g}$ state
to the `6 eV states' of the O$_2$ ${c}^1\Sigma^{-}_{u}$, 
${A'}^3\Delta_{u}$ and ${A}^3\Sigma^{+}_{u}$ states. 
The cross sections for excitation to the individual state are 
shown in thin lines. 
The thick full line represents the sum of these cross sections.  
  }
\end{figure}

%a-> `6 eV states'
Excitation cross sections to the `6 eV states' from the O$_2$ 
${a}^1\Delta_{g}$ state are plotted in figure \ref{fig8}. 
There we show the summed cross sections as well as individual 
contributions of the O$_2$ ${c}^1\Sigma^{-}_{u}$,${A'}^3\Delta_{u}$ 
and ${A}^3\Sigma^{+}_{u}$ states.
The summed total cross section has a similar shape to the excitation 
cross section from the ${X}^3\Sigma^{-}_{g}$ state. 
The origin of a peak at 7.0 eV is the O$_2^{-}$ ${}^{2} \Pi_u$ 
resonance, as in the case of the ${a}^1\Delta_{g}$ $\to$ 
${b}^1\Sigma^{+}_{g}$ transition. 
The magnitude of the cross sections at this resonance peak is about 3 
times larger than the corresponding cross sections of the 
${X}^3\Sigma^{-}_{g}$ case in figure \ref{fig5}. 
The difference is less pronounced in the energy region above 10 eV 
where the cross sections from the ${a}^1\Delta_{g}$ state are about 
30\% larger than those from the ${X}^3\Sigma^{-}_{g}$ state. 

\begin{figure}
 \includegraphics{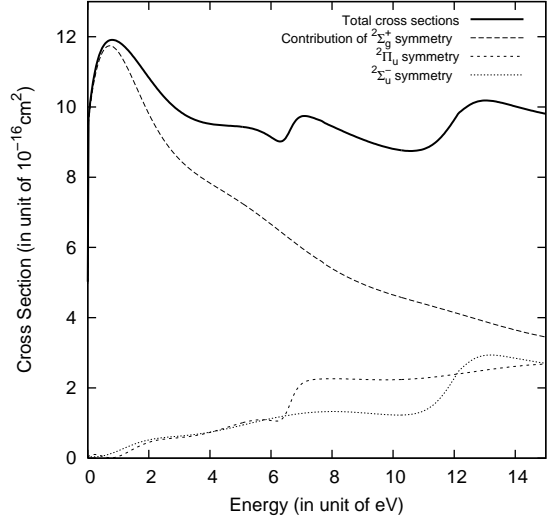}%
 \caption{\label{fig9}
Elastic cross section for the O$_2$ ${b}^1\Sigma^{+}_{g}$ state. 
The results are shown in thick full line with the contributions 
of ${}^{2}\Sigma_{g}^+$, ${}^{2}\Pi_{u}$ and ${}^{2}\Sigma_{u}^-$
symmetries in thin lines as in figure \ref{fig6}.  
   }
\end{figure}

\begin{figure}
  \includegraphics{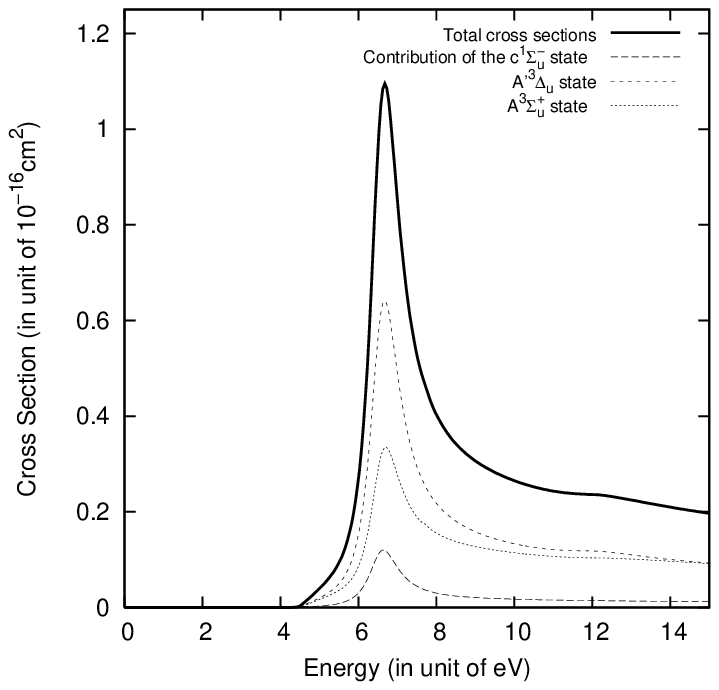}%
  \caption{\label{fig10}
Excitation cross sections from the O$_2$ ${b}^1\Sigma^{+}_{g}$ 
state to the `6 eV states' of the O$_2$ ${c}^1\Sigma^{-}_{u}$, 
${A'}^3\Delta_{u}$ and ${A}^3\Sigma^{+}_{u}$ states. 
The cross sections for excitation to the individual state are 
shown in thin lines. 
The thick full line represents the sum of these cross sections.  
  }
\end{figure}

Figures \ref{fig9} and \ref{fig10} show the cross sections for 
electron collisions with the O$_2$ ${b}^1\Sigma^{+}_{g}$ excited
state.  
The overall features are quite similar to the corresponding cross 
sections from the ${a}^1\Delta_{g}$ state. 
In particular, the elastic cross sections of the ${b}^1\Sigma^{+}_{g}$ 
state in figure \ref{fig9} are almost the same as those of the 
${a}^1\Delta_{g}$ state shown in figure \ref{fig6}. 
Excitation cross sections to the `6 eV states' in figure \ref{fig10} 
are slightly different from the excitation cross sections to the 
`6 eV states' from the ${a}^1\Delta_{g}$ state. 
The height of the cross section peak at 6.5 eV is about 35\% larger 
than that of the ${a}^1\Delta_{g}$ case. 
As in the cases of the ${X}^3\Sigma^{-}_{g}$ $\to$ `6 eV states' and 
${a}^1\Delta_{g}$ $\to$ `6 eV states' excitations, 
the O$_2^{-}$ ${}^{2} \Pi_u$ resonance causes this peak 
in the cross section. 
The location of the peak in figure \ref{fig10} is shifted from the 
peak positions in figures \ref{fig5} and \ref{fig8}   
because of energy differences between the O$_2$ ${X}^3\Sigma^{-}_{g}$,  
${a}^1\Delta_{g}$ and ${b}^1\Sigma^{+}_{g}$ states.  

Finally, we discuss the effect of the extra ${1}^{1,3}\Pi_{g,u}$ 
target states in our R-matrix calculations. 
In figures \ref{fig2}-\ref{fig5}, 
we compare cross sections of electron collisions with 
the O$_2$ ${X}^3\Sigma^{-}_{g}$ state from the 9 target states 
calculations and those from the 13 target states calculations 
including 4 extra ${1}^{1,3}\Pi_{g,u}$ target states. 
Inclusion of extra ${1}^{1,3}\Pi_{g,u}$ target states generally lowers 
the cross sections. 
However, this lowering is less than 15\% and is not
significant. 
Our 9 target states cross sections have similar magnitude in general 
compared to the 9 target-states R-matrix calculations of Noble and 
Burke \cite{No92}. 
However, they are slightly different for the excitations 
to the ${a}^1\Delta_{g}$, ${b}^1\Sigma^{+}_{g}$ states and 
`6 eV states' in the energy region around 8.0 eV. 
In this O$_2^{-}$ ${}^{2} \Pi_u$ resonance region, 
our cross sections are about 20-30\% smaller than their results. 
These differences in the cross sections may be attributed to the 
different treatment of the basis set and the CI representations of 
the target states in ours and their calculations.

\clearpage

\section{Summary}

We have investigated electron collisions with the excited
${a}^1\Delta_{g}$, ${b}^1\Sigma_{g}^+$ states of the O$_2$ molecule 
using the fixed-bond R-matrix method which includes 13 target 
electronic states,
${X}^3\Sigma^{-}_{g}$,${a}^1\Delta_{g}$,
${b}^1\Sigma^{+}_{g}$,${c}^1\Sigma^{-}_{u}$,${A'}^3\Delta_{u}$,
${A}^3\Sigma^{+}_{u}$,${B}^3\Sigma^{-}_{u}$,${1}^1\Delta_{u}$,
${f'}^1\Sigma^{+}_{u}$,
${1}^1\Pi_{g}$,${1}^3\Pi_{g}$,${1}^1\Pi_{u}$ and ${1}^3\Pi_{u}$. 
These target states are described by CI wave functions in the valence 
CAS space, using SA-CASSCF orbitals. Gaussian type orbitals 
are used in this work, in contrast to the STOs in the previous works.   
Our vertical excitation energies are in good agreement the previous 
results and the experimental values. 
We obtaine integral cross sections for 
${a}^1\Delta_{g}$ $\rightarrow$ ${a}^1\Delta_{g}$,${b}^1\Sigma_{g}^+$ 
and `6eV states'(${c}^1\Sigma^{-}_{u}$,
${A'}^3\Delta_{u}$ and ${A}^3\Sigma^{+}_{u}$),  
as well as ${b}^1\Sigma_{g}^+$ $\rightarrow$ ${b}^1\Sigma_{g}^+$ and 
`6eV states'. 
The magnitude of the cross sections for the ${a}^1\Delta_{g}$ 
$\rightarrow$ ${b}^1\Sigma_{g}^+$ transition is consistent with the 
existing experimental value, which is 10 time larger than the one for 
${X}^3\Sigma^{-}_{g}$ $\rightarrow$ ${b}^1\Sigma_{g}^+$. 
The elastic cross sections for the ${a}^1\Delta_{g}$ state and the 
${b}^1\Sigma_{g}^+$ state have similar magnitude and shape when compared 
to the elastic cross sections of the ${X}^3\Sigma^{-}_{g}$ state. 
The transitions for the ${a}^1\Delta_{g}$,${b}^1\Sigma_{g}^+$ 
$\rightarrow$ `6eV states' have cross sections about 5 times larger 
than the corresponding transitions from the ${X}^3\Sigma^{-}_{g}$ 
ground state. 
Our results will be important for modeling of plasma discharge chemistry 
which needs cross sections between the excited electronic states 
in some case.

%%%%%%%%%%%%%%%%%%%%%%%%%%%%%%%%%%%%%%%%%%%%%%%%%%%%%%%%%%%%%%%%%%%%%%%%%%

\begin{acknowledgments}
M.T. thank Dr. Gorfinkiel for her hospitality during his visit to
UCL. 
The present research is supported in part by the grant from the Air 
Force Office of Scientific Research: the Advanced High-Energy 
Closed-Cycle Chemical Lasers project (PI: Wayne C. Solomon, 
University of Illinois, F49620-02-1-0357). 
Computer resources were provided in part by the Air Force 
Office of Scientific Research DURIP grant (FA9550-04-1-0321) as well
as by the Cherry L. Emerson Center for Scientific Computation at 
Emory University. 
The work of M.T. was partially supported by the Japan Society for the 
Promotion of Science Postdoctoral Fellowships for Research Abroad. 
\end{acknowledgments}

\clearpage

%\bibliography{draft}

\clearpage

% If in two-column mode, this environment will change to single-column
% format so that long equations can be displayed. Use
%\begin{widetext}
%\end{widetext}

% figures should be put into the text as floats.
% Use the graphics or graphicx packages (distributed with LaTeX2e)
% and the \includegraphics macro defined in those packages.
% See the LaTeX Graphics Companion by Michel Goosens, Sebastian Rahtz,
% entire page. There is no need to do explicit centering.

% \begin{figure}
% \end{figure}

% Surround figure environment with turnpage environment for landscape
% figure
% \begin{figure}
% \end{figure}

\clearpage

\clearpage

\clearpage

\clearpage

\clearpage

\clearpage

\clearpage

\clearpage

\clearpage

\clearpage

% tables should appear as floats within the text
% The ruledtabular enviroment adds doubled rules to table and sets a
% reasonable default table settings.
% Use the table* environment to get a full-width table in two-column
% Add \usepackage{longtable} and the longtable (or longtable*}
% environment for nicely formatted long tables. Or use the the [H]
% placement option to break a long table (with less control than 
% \begin{table}%[H] add [H] placement to break table across pages
% \begin{ruledtabular}
% Lines of table here ending with \\
% \end{tabular}
% \end{ruledtabular}
% \end{table}

\begin{table}%
\caption{\label{tab0}
Division of the orbital set in each symmetry. 
}
\begin{ruledtabular}
\begin{tabular}{lrrrrrrrr}
Symmetry  & $A_g$ & $B_{2u}$ & $B_{3u}$ & $B_{1g}$ &  $B_{1u}$ &
   $B_{3g}$ &  $B_{2g}$ &   $A_u$    \\
\hline
Valence   & 1-3$a_g$ & 1$b_{2u}$ &  1$b_{3u}$ &  &  1-3$b_{1u}$ &
  1$b_{3g}$  &  1$b_{2g}$ &   \\ 
Extra virtual   & 4$a_g$ & 2$b_{2u}$ &  2$b_{3u}$ & 1$b_{1g}$ & 
4$b_{1u}$ &  2$b_{3g}$ & 2$b_{2g}$  & 1$a_u$  \\
Continuum & 5-38$a_g$ & 3-35$b_{2u}$ &  3-35$b_{3u}$ & 
2-17$b_{1g}$  &  5-37$b_{1u}$ &  3-18$b_{3g}$ &  
3-18$b_{2g}$ & 2-17$a_u$    \\ 
\end{tabular}
\end{ruledtabular}
\end{table}

\clearpage

\begin{table}%
\caption{\label{tab1}
 Comparison of the vertical excitation energies at R=2.3a$_0$ 
from the present CASSCF/GTO calculations with previous work of 
Middleton et al. \cite{Mi94} as well as experimental values quoted in 
Teillet-Billy et al. \cite{Te87}. The unit of energy is eV.
}
\begin{ruledtabular}
\begin{tabular}{lrrr}
State & Present CASSCF/GTO & Previous HF/STO & Experimental values\\
\hline
${X}^3\Sigma^{-}_{g}$ & 0.00 & 0.00 & 0.00 \\
${a}^1\Delta_{g}$     & 0.93 & 0.93 & 0.98 \\
${b}^1\Sigma^{+}_{g}$ & 1.43 & 1.47 & 1.65 \\
${c}^1\Sigma^{-}_{u}$ & 5.60 & 5.49 & 6.12 \\
${A'}^3\Delta_{u}$    & 5.82 & 5.68 & 6.27 \\
${A}^3\Sigma^{+}_{u}$ & 5.93 & 5.81 & 6.47 \\
${B}^3\Sigma^{-}_{u}$ & 9.80 & 10.86 & 9.25  \\
${1}^1\Delta_{u}$      & 12.23 & 13.16 & 11.8  \\
${f'}^1\Sigma^{+}_{u}$  & 13.57 & 14.67 & 13.25 \\
\end{tabular}
\end{ruledtabular}
\end{table}

% Surround table environment with turnpage environment for landscape
% table
% \begin{table}
% \begin{ruledtabular}
% \end{tabular}
% \end{ruledtabular}
% \end{table}

%\appendix
%\section{}

\end{document}